\documentclass[aip,reprint]{revtex4-1}
\usepackage{graphicx}
\begin{document}
\title{Coherent Operation of a Gap-tunable Flux Qubit}
\author{Xiaobo Zhu}
\email[]{xbzhu@will.brl.ntt.co.jp}
\author{Alexander Kemp}
\author{Shiro Saito}
\author{Kouichi Semba}
\email[]{semba@will.brl.ntt.co.jp}
\affiliation{NTT Basic Research Laboratories, 3-1, Morinosato Wakamiya, Atsugi-shi, Kanagawa 243-0198, Japan}
\date{\today}
\begin{abstract}
  We replace the Josephson junction
  defining a three-junction flux qubit's properties with a tunable direct current superconducting quantum interference devices (DC-SQUID) in order to
  tune the qubit gap during the experiment. We observe different gaps
  as a function of the external magnetic pre-biasing field and
  the local magnetic field through the DC-SQUID controlled by high-bandwidth on chip control
  lines. The persistent current and gap behavior correspond to
  numerical simulation results. We set the sensitivity of the gap on the control
  lines during the sample design stage. With a tuning range of several GHz on a qubit dynamics timescale, we observe coherent system dynamics at the degeneracy point.
\end{abstract}
\maketitle
Superconducting flux qubits consisting of three Josephson
junctions embedded in a low inductance superconducting
loop\cite{Mooij99,T.P.Orlando1999} are promising candidates for quantum information
processing\cite{you05}. With the standard design, the magnetic field inducing an
energy difference between the ground state and the first excited state is the {\em only parameter that is adjustable during the experiment}. At the degeneracy point,
it reaches its minimum value (called the gap $\Delta$), which is related to the quantum mechanical
tunnel rate. However, the gap is determined at the time of production by the inherent
parameters of the three Josephson junctions.
One of the main problems in realizing quantum computation schemes using flux
qubits is 1/f flux noise. At the degeneracy point the qubit is
decoupled from low frequency flux
noise\cite{Yoshihara06,Kakuyanagi07}, and achieves maximal coherence
times of several $\mu{}s$ rather than $ns$ at off-degeneracy
point, wherefore this is the optimal operation point.
Implementing a quantum processor based on flux qubits requires the
operation at this optimal point at all times, especially when coupling
several qubits.  Recently, circuit quantum electrodynamics (QED) (quantum bus, qubus) schemes have been
developed for many superconducting
qubits\cite{Wallraff04,Chiorescu04,Leek07,DiCarlo09,Johansson06}, however standard
flux qubits suffer from the fact that bringing them in resonance with
such a quantum bus requires to operate them at the off-degeneracy point.

With the flux qubit we present here we aim to overcome these limitations by
replacing the smallest junction of the flux qubit with a low
inductance DC-SQUID loop\cite{Mooij99,T.P.Orlando1999}. Varying
the magnetic flux penetrating this loop is equivalent to changing the
critical current of the smallest junction, thus making the gap tunable during the experiment. Using this
tuning parameter we can control the coupling to a qubus by tuning into
or out of the qubus frequency while operating at the optimal
point. Moreover, we can implement strong transverse coupling to another qubit or
to a resonator\cite{Wang09,Kerman08}.
\begin{figure}
\includegraphics{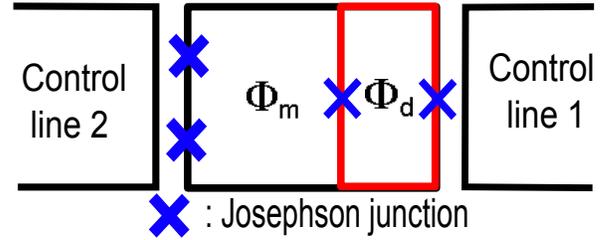}%
\caption{ The schematic diagram of our qubit design showing the main
  loop (left loop), the DC-SQUID loop (right red loop) and the control lines.}
  \label{1}%
 \end{figure}

 Our qubit is shown schematically in Fig.~1. Two identical
 junctions with Josephson energy $E_J$ in series with a symmetric
 DC-SQUID, in which each junction has a Josephson energy of $\alpha_0
 E_J $ are enclosed in a superconducting loop. The effective Josephson
 energy of the DC-SQUID is $\alpha E_J=2 \alpha_0 \cos(\pi\Phi_d/\Phi_0)
 E_J$, where $\Phi_d$ is the flux threading the DC-SQUID, and $\Phi_0$
 is the flux quantum. The main loop is threaded by a flux
 $\Phi_m$. The phase drop caused by $\Phi_d$ adds up to an effective
 magnetic flux $\Phi_t = \Phi_m+\Phi_d /2$ threading the qubit.  We
 operate at fluxes close to $\Phi_t = \Phi_0 (n+1/2)$, where $n$ is an
 integer. At these points the system acts as a qubit with the
 Hamiltonian
 \begin{equation}
   H=\frac{\varepsilon(\Phi_t)}{2}\sigma_z+\frac{\Delta(\alpha)}{2}\sigma_x,
 \end{equation}
 where $\varepsilon=2I_p(\Phi_t-\Phi_0 (n+1/2) )$ is energy difference
 between the two different classical persistent current states induced by
 the external magnetic field, and $\Delta(\alpha) \equiv \Delta(2 \alpha_0
 \cos(\pi\Phi_d/\Phi_0))$ is the tunnel element between these two
 states, and $\sigma_x$ and $\sigma_z$ are Pauli matrices.

There is intrinsic crosstalk between the DC-SQUID flux and the
effective qubit flux, unrelated to an inductive crosstalk of the control lines to either loops, since $\Phi_t=\Phi_m+\Phi_d /2$. This strong crosstalk will
shift the operation point of the qubit, thus causing strong dephasing and transtions to higher levels. F.~G.~Paauw \emph{et al.}\cite{Paauw09} utilize a
gradiometric double loop design to solve this problem. With N fluxoids (N is an odd integer) trapped in the superconducting loop of the gradiometer the qubit is pre-biased closed to its degeneracy point. They realized the \emph{in situ} tunability of a gap. However,
 a short relaxation time prevented further experiments\cite{note1}.

In our design, we place two control lines, 1 and 2 (Fig. 1), adjacent to the qubit structure. The fluxes in the qubit loops are related to the control currents by the mutual
 inductances to the respective lines. The mutual inductances of control
line 1 to the DC-SQUID loop and main loop are 85 fH and 64 fH, and
those of control line 2 are 1 fH and 84 fH respectively. We carry out our experiment as follows: first we pre-bias the effective qubit flux $\Phi_t$ close to the
operating point $(n+1/2)\Phi_0$ by using the external magnetic coil located in
the Dewar. This operation is slow, since the
solenoid has a settling time on the order of minutes. Applying currents to the control lines tunes $\Phi_t$ and $\Phi_d$ \emph{in situ} away from their pre-biased values. The
coupling to the effective qubit loop induced by control line 1 can be
compensated by applying the corresponding current to control line 2,
thus giving full control over the two-dimensional parameter space.
\begin{figure}
\includegraphics{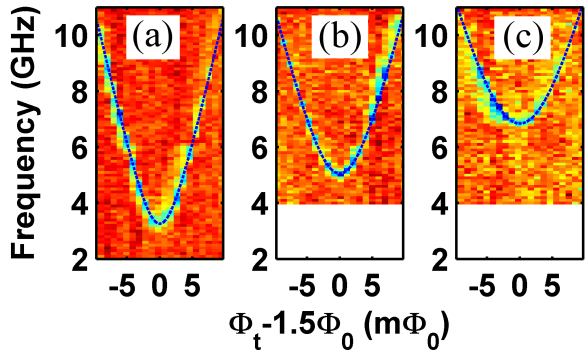}%
 \caption{\label{2} Spectra with different gaps: the gaps $\Delta$ and persistent currents $I_p$ are
 (3.26 GHz,166.8 nA)(a), (5.05 GHz, 155.3 nA)(b) and (6.84 GHz, 144.6 nA)(c) by setting shift pulse voltage level
 of control line 1 at 0.0, 1.2 and 2 V respectively. $\Delta$ and $I_p$ were obtained by fitting the spectra to Eq. 2 (blue dashed line) }%
 \end{figure}

 In Fig. 2(a-c), we show the change in the
 gap $\Delta$ as a function of the pulse voltage level of control line 1
 while using control line 2 to scan the parameter
 effective magnetic flux $\Phi_t$. In this measurement, $\Phi_t$ is pre-biased to an operation point near $3/2\Phi_0$ (See below for details on the choice of the operation point). Because the mutual inductance of the
 control line 2 to the DC-SQUID loop is much smaller than that to the
 main loop, we can neglect the influence of control line 2 on
 $\Phi_d$ and assume that the gap is constant during each scan. Therefore, the spectra obtained here are similar to that of a traditional 3JJ flux qubit. Each curve can fit to the dispersion relation of the
 qubit Hamiltonian in Eq.~1, given by
 \begin{equation}\mathit{\Delta} E=\sqrt{\varepsilon(\Phi_t)^{2}+\Delta(\alpha)^2},
 \end{equation}
 so we obtain the gap $\Delta$ and the persistent currents $I_p$ by
 fitting.

 The bandwidth of our
 control lines is 20 GHz. This imposes no relevant limit to the proposed
 schemes\cite{Wang09} on the speed with which we can tune the flux of
 the loops \emph{in situ}. On the other hand, because we now intend to tune the qubit
 adiabatically, the tuning frequency should be much smaller than the
 energy level space between the ground state and the first excited
 state. Hence, we limited the pulse rise-and-fall times of the pulse generators in
 this measurement to 1.6 ns. This value is also suitable for
 demonstrating selective qubit coupling to a quantum bus by tuning
 into resonance.

\begin{figure}
\includegraphics{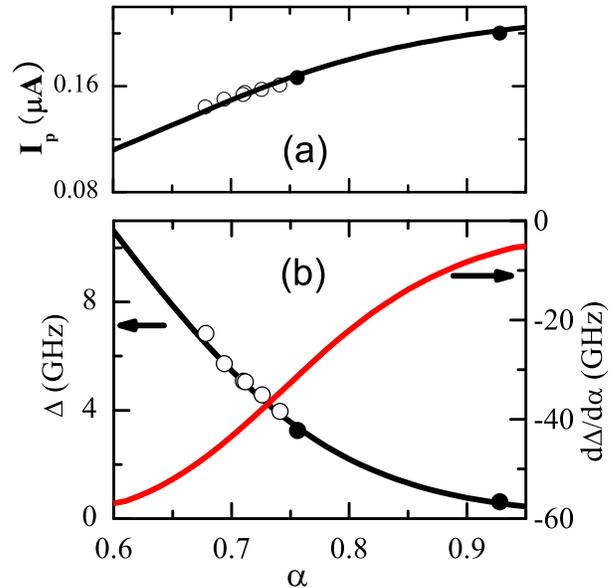}%
 \caption{\label{3} The persistent currents $I_p$ (a) and the gap (left y axis) and its derivative (red line, right y axis) (b)
as a function of the $\alpha$ values. The circles are the experimental data and lines are the fits.
$\Phi_d$ can be changed by the external magnetic coil and control line 1, but for two of them (solid circles) we do not add flux via control line 1.
}%
\end{figure}

We extract the persistent currents and the gaps from the spectra, and plot them in Fig.~3 (circles) as a function of
$\alpha$ values (three of the original spectra are shown in Fig.~2). The lines are fitting results of the numerical diagonalization of the full discretized Hamiltonian\cite{T.P.Orlando1999} in a finite difference scheme. The charging energy $E_c$=5.05 GHz, the Josephson energy $E_J$=139 GHz and $2\alpha_0$=0.95 in the
numerical calculation  are obtained by fitting to the experimental data. The charging energy $E_c$ compares well with the values obtained earlier for similar production
parameters\cite{Deppe04}.

It should be pointed out that the sensitivity of the gap on the DC-SQUID loop frustration
$|d\Delta/d\Phi_d|$=$|(d\Delta/d\alpha)\cdot (d\alpha/d\Phi_d)|$ depends on the pre-biasing point. When $\Phi_t$ is pre-biased to $(n+1/2)\Phi_0$ and no shift pulse is added to the qubit, $\Phi_d$ is only determined by the external magnetic coil and is equal to $(n+1/2)\gamma\Phi_0$, where the factor
$\gamma$=$S_d$/$S_t=0.138$ is the ratio of the DC-SQUID loop area $S_d$ to
the effective qubit loop size $S_t$ which is $S_t=S_m+1/2 S_d$.
Since $\frac{d\alpha}{d\Phi_d}=\frac{2\pi\alpha_0}{\Phi_0}\sin(\pi(n+1/2)\gamma)$, and $d\Delta/d\alpha$ obtained from Fig.~3(b), we can calculate the $|d\Delta/d\Phi_d|$ value at each pre-biasing point.
At the working points selected in this measurement, $|d\Delta/d\Phi_d|$ is equal to
3.96 GHz/$\Phi_0$ (at $\Phi_d=1/2\gamma\Phi_0$ ) and 56.5 GHz/$\Phi_0$
(at $\Phi_d=3/2\gamma\Phi_0$) (these two points are shown by black solid circles in Fig.~3). This calculation shows
that the sensitivity at $\Phi_d=3/2\gamma\Phi_0$ is more than ten times
larger than that at $\Phi_d=1/2\gamma\Phi_0$. So, we choose the bias
point $\Phi_t=3/2\Phi_0$ as the working point
for experiments on controlling the gap (open circles in Fig.~3). We can control the sensitivity at
a given qubit gap during the design of the sample by adjusting
$\alpha_0$, $\gamma$ and choosing the pre-bias point appropriately to obtain a larger sensitivity. However, the greater sensitivity also increases the effect of the flux
noise. Therefore, future applications of these kind of
gap-tunable flux qubits require to balance these two conflicting factors.

\begin{figure}
\includegraphics{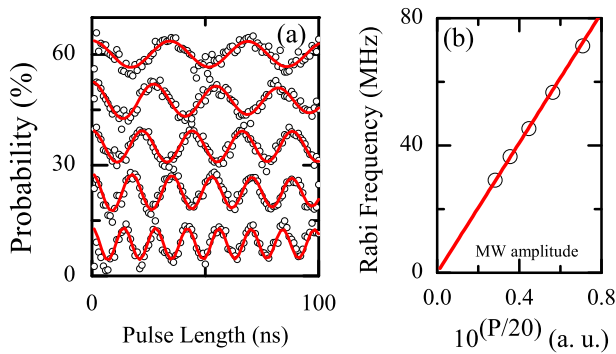}%
 \caption{\label{4} (a) Rabi oscillations at one degeneracy point for five different microwave
powers (open circles). The data are fitted by exponentially damped sinusoidal oscillations (red line). (b) Linear dependence
of the Rabi frequency on the microwave amplitude.}%
 \end{figure}
 Rabi oscillations between the ground state and
 the first excited state are an elementary demonstration of a single qubit
 coherent operation. In Fig.~4(a), we show Rabi oscillations obtained at the
 degeneracy point. We set the voltage
level of the two control lines to the degeneracy point
of a dispersion relation, and varied the length of the microwave pulse
 resonant with the qubit, i.e. the gap frequency. After above sequence, the qubit state is measured by
applying a measurement pulse\cite{Johansson06}. We repeated the whole sequence 2000 times to detect the relative occupation of the ground state and the
 excited state. We varied the amplitude of the microwave, and verified
 the linear dependence of the Rabi frequency on the microwave
 amplitude as shown in Fig.~4(b), which is a signature of the Rabi process.

 In summary, we designed and fabricated an improved flux qubit, whose gap can be tuned \emph{in situ} by
 two high bandwidth control lines. We also demonstrated coherent
 Rabi oscillations at the degeneracy point. This gap tunable flux qubit is
 a promising candidate for the implementation of scalable quantum
 computation.
\begin{acknowledgments}
We would like to thank H. Nakano, K. Kakuyanagi, Y. D. Wang and S. Karimoto for fruitful discussions. This work was supported in part by Funding Program for World-Leading
Innovative R\&D on Science and Technology(FIRST), Scientific Research of
Specially Promoted Research 18001002 by MEXT, and Grant-in-Aid for Scientific Research (A) 18201018 and 22241025 by JSPS.

\end{acknowledgments}
%
%

\end{document}